\documentclass{aastex6}

\usepackage{graphicx}

\begin{document}

\title{VVV Survey Microlensing Events in the Galactic Center Region
}

\author{
Mar\'ia Gabriela Navarro\altaffilmark{1,2},
Dante Minniti\altaffilmark{1,2,3},
Rodrigo Contreras Ramos\altaffilmark{2,4}
}

\altaffiltext{1}{Depto. de Cs. F\'isicas, Facultad de Ciencias Exactas, Universidad Andr\'es Bello, Av. Fernandez Concha 700, Las Condes, Santiago, Chile.}
\altaffiltext{2}{Millennium Institute of Astrophysics, Av. Vicuna Mackenna 4860, 782-0436, Santiago, Chile.}
\altaffiltext{3}{Vatican Observatory, V00120 Vatican City State, Italy.}
\altaffiltext{4}{Instituto de Astrof\'isica, Pontificia Universidad Cat\'olica de Chile, Av. Vicuna Mackenna 4860, 782-0436 Macul, Santiago, Chile.}

\begin{abstract}
We search for microlensing events in the highly reddened areas surrounding the Galactic center  
using the near-IR observations with the VISTA Variables in the V\'ia L\'actea Survey (VVV).
We report the discovery of 182 new microlensing events, based on observations acquired between the years 2010 and 2015.   
We  present the color-magnitude diagrams of the microlensing sources for the VVV tiles b332, b333, and b334, which were independently analyzed, and show good qualitative agreement amongst themselves. We detect an excess of microlensing events in the central tile b333  in comparison with the other two tiles, suggesting that the microlensing optical depth keeps rising all the way to the Galactic center.  
We derive the Einstein radius crossing time for all of the observed events. The observed event timescales range from $t_E= 5$ to $200$ days.
The resulting timescale distribution shows a mean timescale of $<t_E>=30.91$ days for the complete sample ($N=182$ events), and $<t_E>=29.93$ days if restricted only for the red clump (RC) giant sources ($N=96$ RC events). There are 20 long timescale events ($t_E \geq 100$ days) that suggests the presence of massive lenses (black holes) or disk-disk event. 
This work demonstrates that the VVV Survey is a  powerful tool to detect intermediate/long timescale microlensing events in highly reddened areas, and it enables a number of future applications, from analyzing individual events to computing the statistics for the inner Galactic mass and kinematic distributions, in aid of future ground- and space-based experiments.
\end{abstract}

\section{Introduction} 
\label{sec:intro}

The idea proposed by \cite{Paczynski86}, based on the works of \cite{Einstein16,Einstein36}, that microlensing events can be detected by measuring the intensity variations of millions of stars was highly successful. In particular, the main groups dedicated to observe the Galactic bulge like the Massive Astrophysical Compact Halo Objects (MACHO; \citealt{Alcock93}), the Optical Gravitational Lensing Experiment (OGLE; \citealt{Udalski93}), the Microlensing Observations in Astrophysics (MOA; \citealt{Bond01}), the Expérience pour la Recherche d?Objets Sombres (EROS; \citealt{Aubourg93}), the Disk Unseen Objects (DUO; \citealt{Alard95}), the Wide-field Infrared Survey Explorer (WiSE; \citealt{Shvartzvald12})
and the Korea Microlensing Telescope Network (KMTNet; \citealt{Kim10}, \citealt{kim17}), discovered thousands of events to date in the bulge. These are all optical surveys, and necessarily monitored the regions with low relative extinctions toward the bulge. The innermost regions close to the Galactic center, which are not only severely crowded, but also heavily obscured by interstellar dust, have remained hidden for microlensing up to now. However, these regions are very interesting because this is where we expect to find the highest number of microlensing events and presumably also the largest microlensing optical depth because of the high density of stars \citep{Gould 95}.

Fortunately, in the near-IR, we can penetrate through the gas and dust in this region in order to detect microlensing events. The first such near-IR study was successfully carried out recently by \cite{Shvartzvald17}, who found five highly extinguished microlensing events between 1 and 2 degrees from the Galactic center. 
The \emph{VISTA Variables in the V\'ia L\'actea Survey} (VVV; \citealt{Minniti10}) is a near-IR variability Survey that scans 560 square degrees in the inner Milky Way  using the  \emph{Visible and Infrared Survey Telescope for Astronomy } (VISTA), a 4 m telescope located at ESO's Cerro Paranal Observatory in Chile.  
The main goal of the VVV survey is to create a 3D map of the inner Galaxy, mainly using the $K_s$-band to search for variable stars as distance indicators and tracers of stellar populations.
At the same time, the VVV survey is an excellent tool to detect microlensing events. Even though the VVV survey cadence (nightly at best) is inadequate to routinely detect objects associated with short timescales that should be numerous in the Galactic center region \citep{Gould 95}, this is sufficient to perform a census of  microlensing events toward the central most  part of the Galaxy. 

The analysis of a complete a sample of microlensing events in the central part of the Galaxy has many applications, ranging from the study of the most interesting isolated events: for example, the ones that have long durations which statistically favor more massive lenses to large statistical studies of Galactic structure and evolution. For the latter, the distribution of timescales can be useful to test the different possible scenarios for the structure and evolution of the inner part of the Galaxy (\citealt{Calchi08}, \citealt{Sumi13}). We note that as the timescale is a degenerate combination of lens mass, and lens-source relative parallax and proper motion, it is necessary to include Galactic models related to specific populations.
Moreover, the study of the event rate can be extremely useful to optimize the observational campaign for the Wide Field Infrared Space Telescope (WFIRST) (\citealt{Green12}, \citealt{Spergel15}), and as complementary to the pioneering work published by \cite{Shvartzvald17}. 

The purpose of this paper is to present the first large sample of microlensing events in the Galactic center area using the VVV data. In this work, we use the simple model of lensing by an isolated point mass (PSPL). We derive the Einstein radius crossing time distribution of the observed events.  
We also characterize the microlensing sources using the available near-IR photometry.
For the future, we propose to extend the spatial and temporal range of the sample to compare the observed distributions with the most recent Galactic models and to analyze selected events. 

In \autoref{sec:sec2} we describe the data used in this research and the procedure that was carried out to detect the microlensing events. The characterization of the final sample is shown in \autoref{sec:sec3}. Finally, the conclusions are presented in \autoref{sec:sec5}. \\

\section{THE SEARCH FOR MICROLENSING AROUND THE GALACTIC CENTER}
\label{sec:sec2}
The VISTA telescope is equipped with the \emph{Wide-field VISTA InfraRed Camera} (VIRCAM; \cite{emerson10}) containing 67 million pixels (16 chips of 2048x2048 pixels). The Field of View (FoV) is  $1.501 deg^2$, which is called a ``tile''. The entire VVV observations comprise  196 tiles in the bulge and 152 in the disk area \citep{Saito12}.  The VVV observational schedule includes single-epoch photometry in $ZYJHK_s$ bands and variability campaign in $K_s$ band \citep{Minniti10}. 
In this work, we focus on the innermost tiles of the VVV (b332, b333 and b334), where the crowding is so severe that PSF photometry is mandatory. Accordingly, the photometric reduction of each detector was carried out using the DAOPHOT II/ALLSTAR package \citep{stetson87}, and the catalogs made at the Cambridge Astronomical Survey Unit (CASU) with the VIRCAM pipeline v1.3 \citep{irwin04}
 were used to calibrate our photometry into the VISTA system by means of a simple magnitude shift using several thousands stars in common (see \citealt{contreras17}). We specifically applied this procedure separately on each detector of the tiles b332, b333, and b334 located within $1.68^o \geq l \geq -2.68^o$ and $ 0.65^o \geq b \geq -0.46^o$ in the Galactic bulge. We detected a total of approximately $14 \times 10^6$ point sources in these three tiles, for which multi-epoch magnitudes in the $K_s$-band were measured. The reduced data included about 100 epochs spanning six seasons (2010-2015) of observations. 

The search of events was performed by means of a new reduction code specially developed for microlensing detections. Contrary to the classical variable star detection, our approach has been optimized to keep those events showing a few deviating points with a transient magnification of the apparent brightness, which would be likely rejected using the classical variability indexes.
This procedure delivers a quality index for each light curve related to how similar it is with a microlensing curve. It is then necessary to cull the sample by selecting the curves with higher quality indices for subsequent visual inspection, but before that it is crucial to perform the fitting procedure using the simplest model, assuming a point source and a point lens (PSPL) \citep{refsdal64}.  Where $F = F_s A(u(t))$, with $F$ being the observed and $F_s$ the catalog source flux, for their non-blended fits. The amplification $A(u(t))$ and the angular distance between the lens and the source projected on the plane of the lens in Einstein radii units $u(t)$ can be written as \\

\noindent\begin{minipage}{.5\linewidth}
\begin{equation}
A(u(t)) = \frac{u^{2} +2}{u \sqrt{u^{2} +4}} 
\end{equation}
\end{minipage}%
\begin{minipage}{.5\linewidth}
\begin{equation}
u(t) = \sqrt{u_{0}^2 + \left( \frac{t-t_{0}}{t_E} \right)^2}
\end{equation}
\end{minipage}

\begin{figure}[t]
\begin{center}
\includegraphics[height = 10 cm]{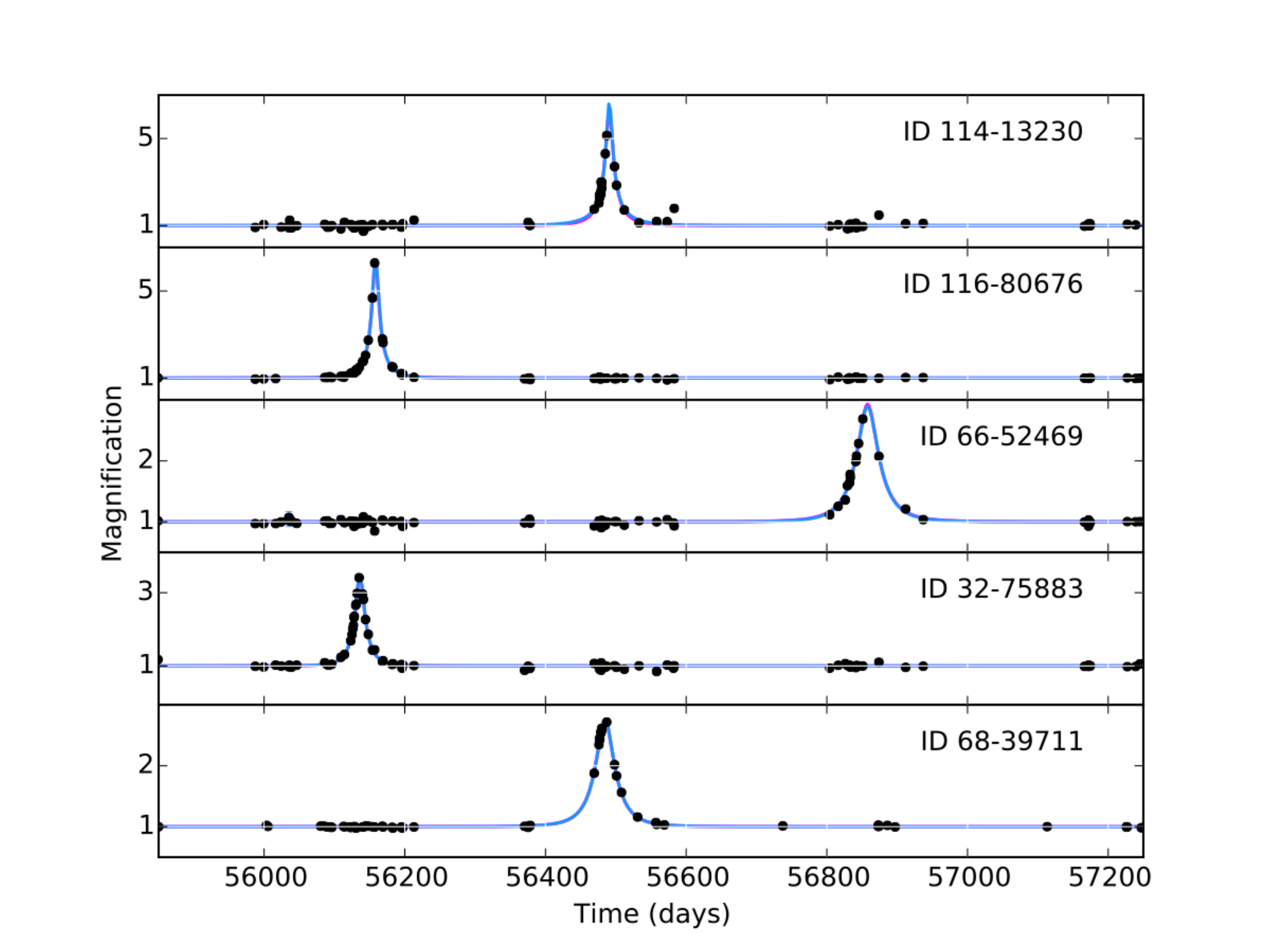}\\
\caption{Sample microlensing light curves and best fits. The first four events indicated in the upper panel are located in the Red Clump.
The fits with (blue line) and without blending (magenta) are indistinguishable and overlap with each other, yielding similar parameters.}
\label{lc}
\end{center}
\end{figure}

The standard microlensing model delivers the $u_0$ related to the impact parameter and thus with the amplitude of the light curve, the time of maximum amplification $t_{0}$ and the Einstein radius crossing time $t_E$. The fitting procedure was performed twice, also including the blending parameter $f_{bl}$ which we expect to be non-negligible in this region, in this case $F =  F_s [f_{bl}(A(u) - 1) + 1]$. \autoref{lc} shows five examples of our near-IR microlensing light curve fits. In all cases, consistent results were found using both procedures. 

At the visual inspection stage, the following requirements were applied for the curve to be qualified as a microlensing event:

\begin{enumerate}
\itemsep-0.5em 
\item Constant baseline;
\item Baseline covering more than one season;
\item At least four points with 4$\sigma$ above the baseline;
\item At least one data point in the rising and falling microlensing light curve;
\item Symmetry during the event;
\item Timescales within an acceptable range to avoid confusions with long period variable stars; and
\item Good fit to single microlensing curve.
\end{enumerate}

The final sample was divided in two groups. The 182 first quality microlensing events that satisfy all the requirements mentioned above (\autoref{tab:table1}), and a second quality list with events showing an evident microlensing light curve, but not meeting all the requirements listed above. We also notice that the last condition eliminated a few good candidate binary events. Hereafter, we will only deal with the high-quality sample, and the individual study of these other cases is deferred for the future.

\begin{figure}[t]
\begin{center}
\includegraphics[height = 6 cm]{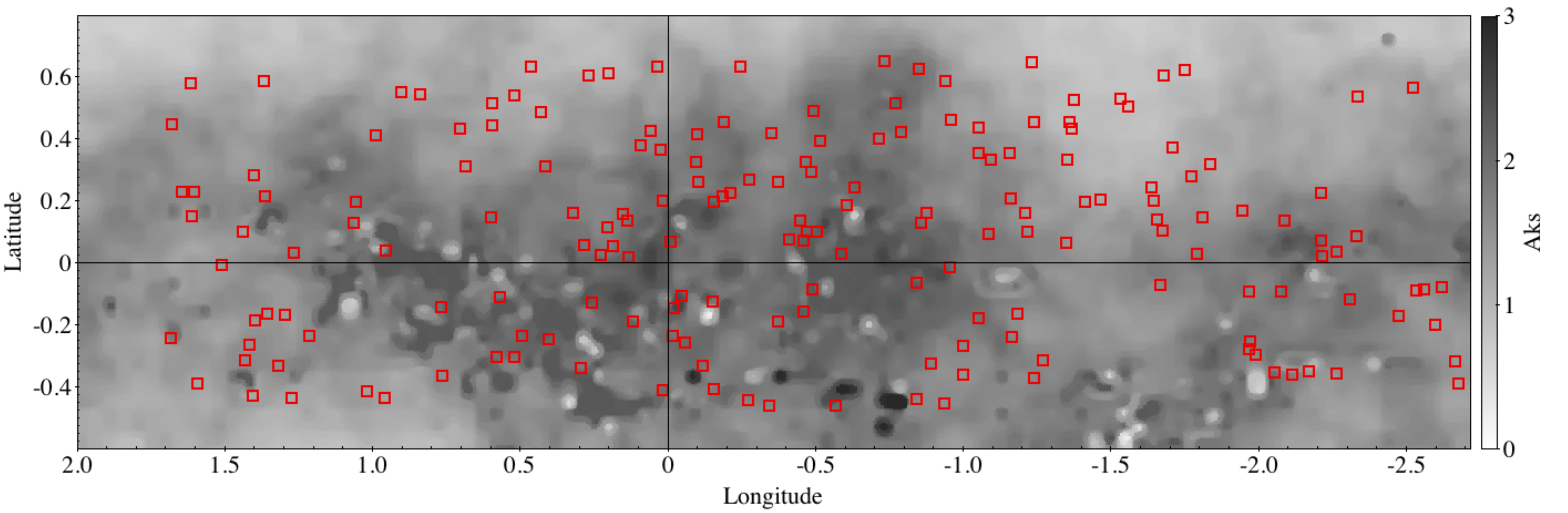}\\
\caption{Spatial distribution of the new microlensing events (red squares) around the Galactic center, overlaid on the extinction map of \cite{gonzalez12}. The duplicate events in the overlapping areas have been accounted for.}
\label{map}
\end{center}
\end{figure}

The magnitude range of the majority of  bulge source stars is $11< K_s <17.5$, and their near-IR colors ($2 < J-K_s < 7$ mags) confirm that they are heavily reddened objects, consistent with most of them being  located in the vicinity of the Galactic center.
The spatial distribution of the final sample of microlensing events is shown in \autoref{map}, where it can be appreciated that we detect events as close as $10$ arcmin from the Galactic center. The distribution  is homogeneous in general, with certain small spatial gaps, which can be attributed in some cases to an increase in the differential reddening.  There are a few over densities that do not appear to be statistically significant. However, the  tile b333 containing the Galactic center has more events than the other two tiles on average. 

Even though tile b333 is the most reddened and crowded of all? and therefore it should be the most incomplete? there is a significant excess of microlensing sources in this central tile.
This is evident if we count only the bright sources with $K_s<16$, where there are $N=78$ sources in tile b333 versus $N=45$ on the average of tiles b332 and b334. This is also seen if we count only the RC sources, where the counts are $N=37$ for b333 versus $N=30$ for the average of the other two tiles, but this is not statistically significant. The most straightforward implication is that the microlensing optical depth keeps rising all the way to the Galactic center, but further observations are necessary to confirm this.

As an external check on the fidelity of our results, we performed the microlensing search separately in the three VVV tiles b332, b333, and b334. There is a small observed overlap region between these tiles, and although the area of the superposition is small ($\sim 4\%$), there is a non-zero probability that the same microlensing event can be detected twice as separate events. To evaluate these cases, we analyzed the events that fulfilled the following conditions simultaneously: distance difference less than 2 arcsec, difference between the time of maximum amplification $t_{0}$ less than 7 days, and  difference between the baseline magnitudes less than 0.15 mag. 
We detected six repetitions in total, and in all these cases we obtained consistent results:
the positions RA and DEC repeat to better than 1 arcsec, 
the $K_s$-band magnitudes repeat to better than 0.08 mag, 
the times of maxima repeat to better than 3 days, and 
the timescales repeat to better than 15\% in all cases but two (these are two short timescale sources  that have a timescale difference of 25\%). 
For these objects, the fitting procedure using the standard microlensing model was recalculated using the data by joining both independent light curves in order to obtain more precise parameters. 

Other checks were made, such as analyzing the timescale versus amplitude relation (\autoref{fig:timescale}). This showed a homogeneous distribution of the amplitudes and no trends with the timescales, as expected. Also, we fitted known microlensing events from OGLE and MOA in order to confirm that our fitting routines yield the correct parameters.

\section{CHARACTERIZATION OF THE MICROLENSING EVENTS}
\label{sec:sec3}

The most important parameter that the standard microlensing model fit provides is the Einstein radius crossing time $t_E$ which is related to the mass of the lens. The precise value of the lens mass can be constrained with the timescale obtained from the light curve; relative distances between the observer; lens and source; and transverse velocity.

\begin{figure}
\centering
\begin{tabular}[b]{@{}p{0.25\textwidth}@{}}
\centering\includegraphics[width=1.3\linewidth]{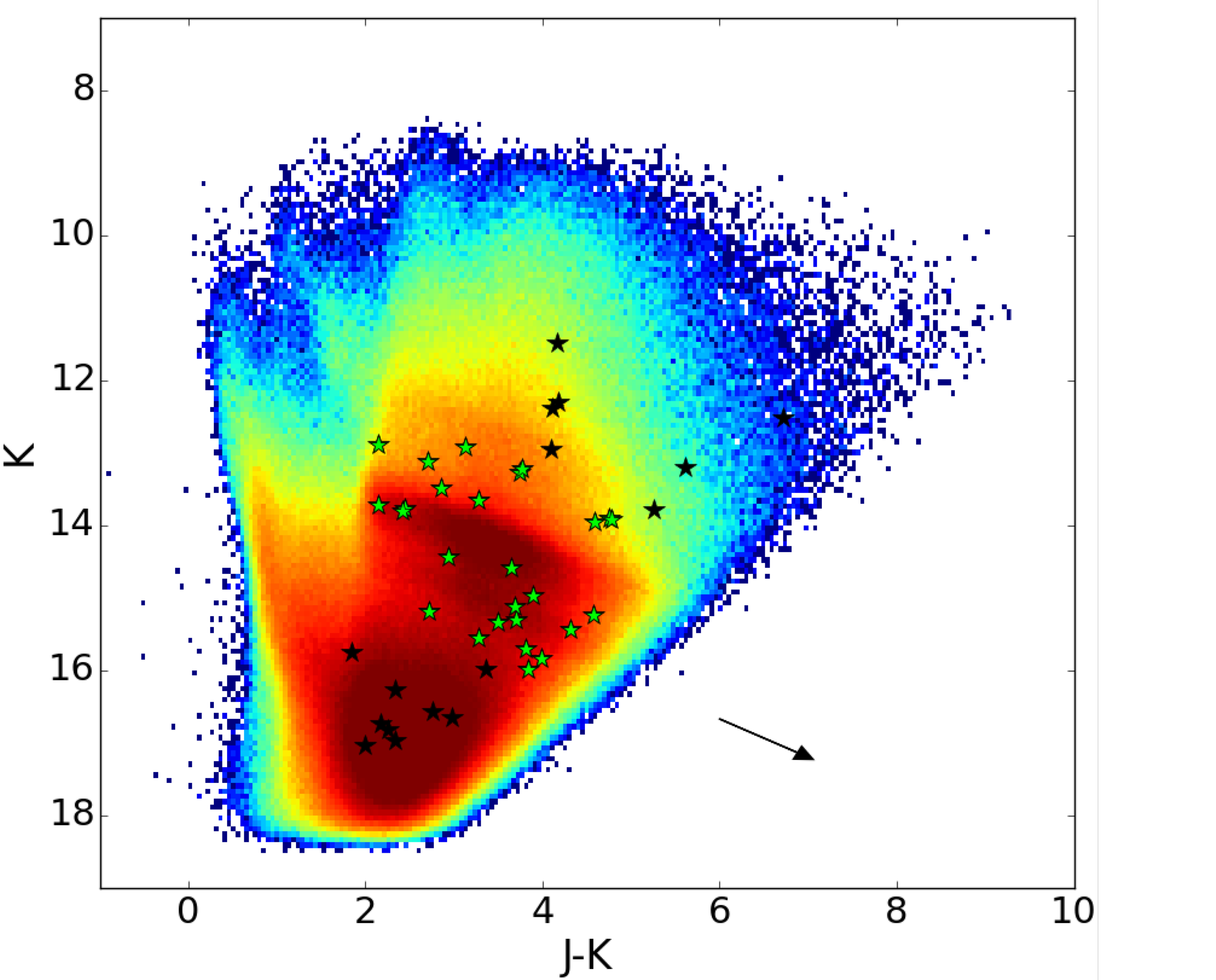} \\
\end{tabular}
\quad
\begin{tabular}[b]{@{}p{0.25\textwidth}@{}}
\centering\includegraphics[width=1.3\linewidth]{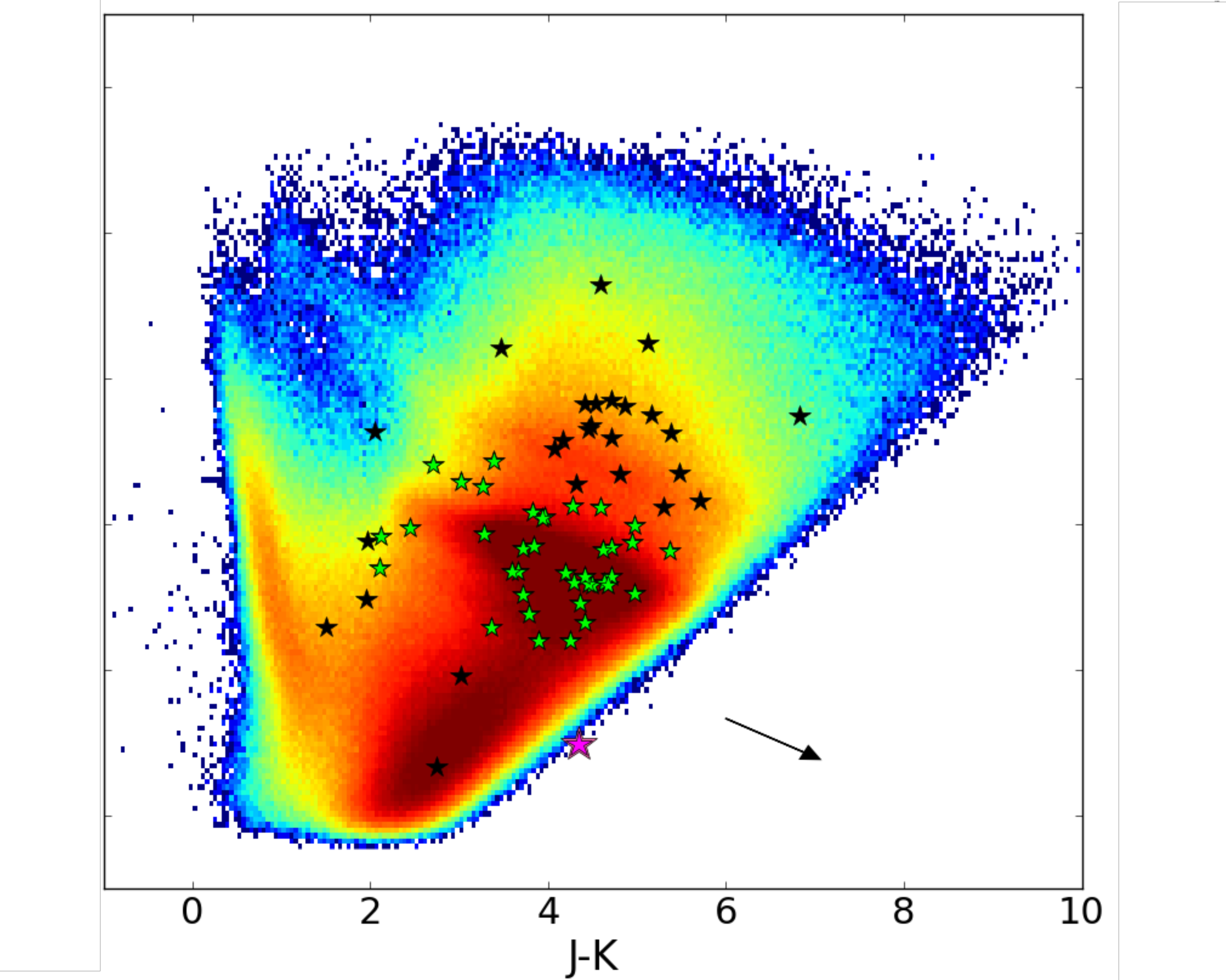} \\
\end{tabular}
\quad
\begin{tabular}[b]{@{}p{0.25\textwidth}@{}}
\centering\includegraphics[width=1.3\linewidth]{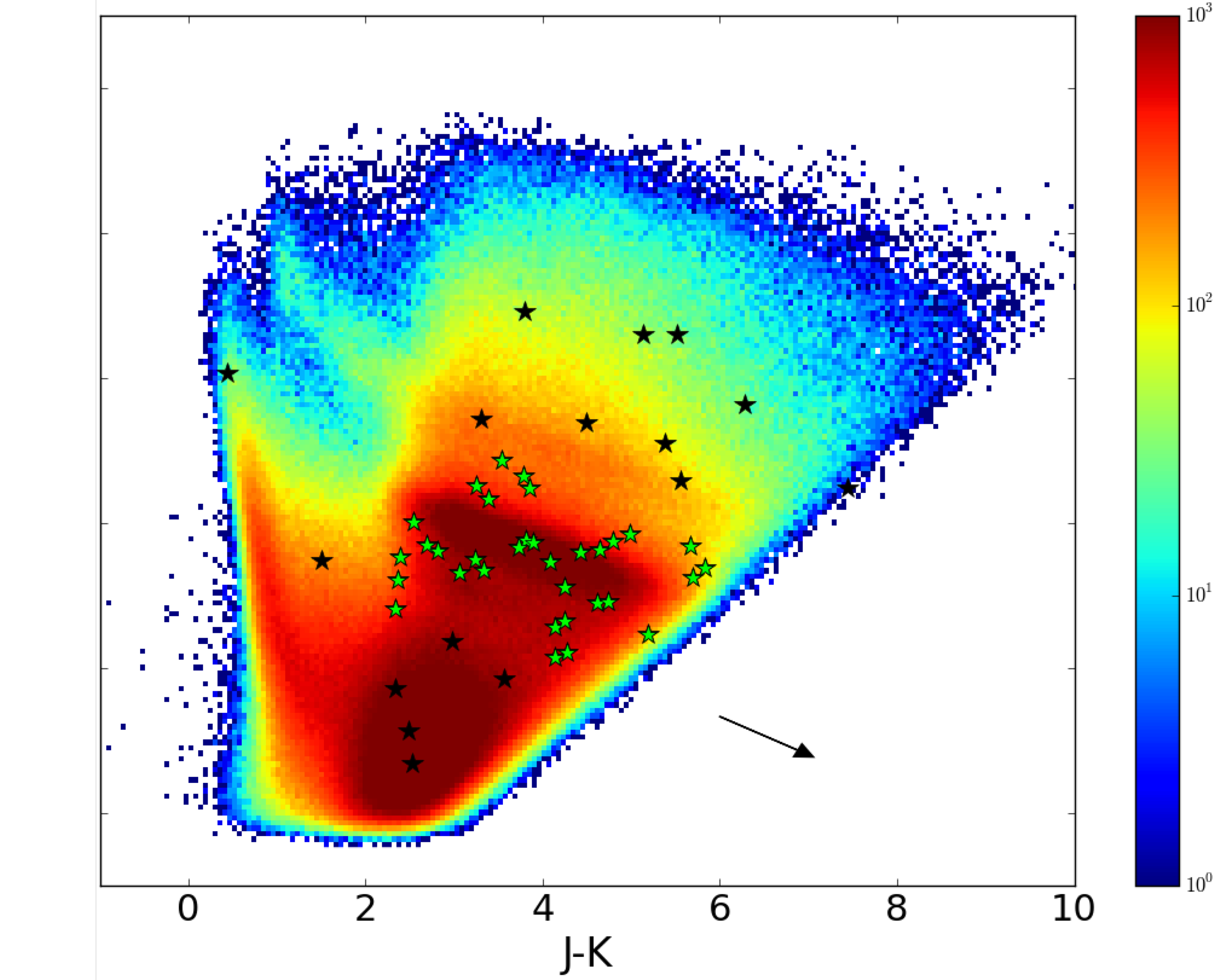} \\
\end{tabular} 
\caption{Near-IR $K_s$ $vs$ $J-K_s$ color-magnitude diagrams for the VVV tiles 332 (left), 333 (center), and 334 (right). The stars indicate the sources of the sample microlensing events. The stars in green are the microlensing events with RC sources. The magenta star in the 333 CMD corresponds to the event with $J$ mag above the detection limit. The arrows show the reddening vector after \citep{nishiyama09}.}
\label{fig:cmd}
\end{figure}

RC giants are core-He burning giants that have known mean luminosities and can be used as distance indicators. Therefore, selecting RC stars with the correct magnitudes increases the probability that they are located at the bulge distance (e.g. \citealt{Popowski05}).
We therefore selected a subsample of events consistent with RC by making magnitude cuts in the color-magnitude diagrams that follow the direction of the reddening vector (\autoref{fig:cmd}). For these RC sources,
we can  assume that they are located in the Galactic bulge. The large reddening is evident, especially in the central most region (tile b333).
Moreover, as blending can be severe in the area we analyzed, the sources that belong to the RC are brighter and give us more reliable information, reducing the blending problem (\citealt{Popowski05}, \citealt{Sumi16}). 
From the color-magnitude diagrams of \autoref{fig:cmd}, it is clear that nearly half of the sources are located in the RC. As a consistency check, all three tiles investigated independently (b332, b333, and b334) show good agreement with each other.

\begin{figure}
\centering
\begin{tabular}[b]{@{}p{0.425\textwidth}@{}}
\centering\includegraphics[width=1.2\linewidth]{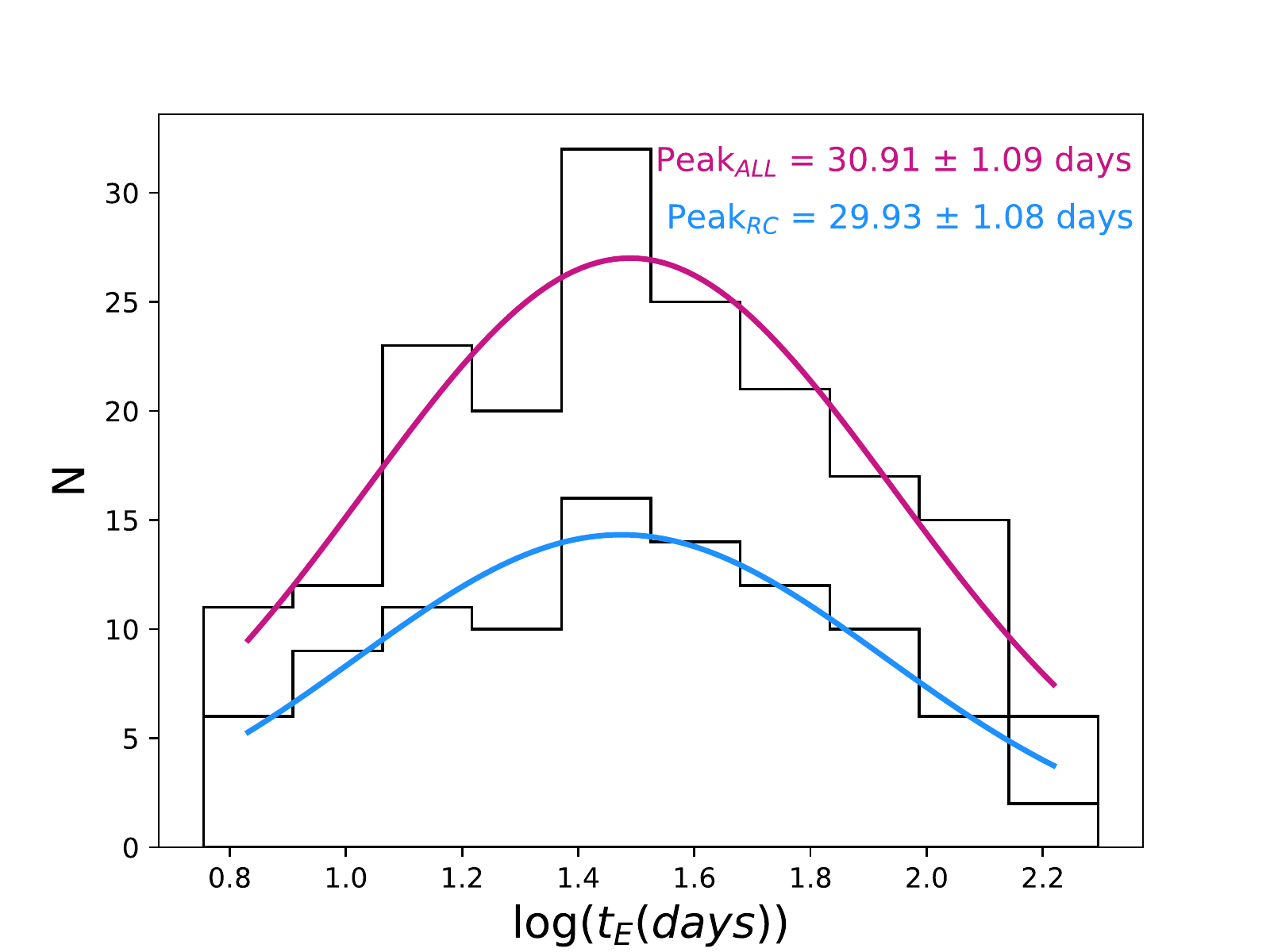} \\
\end{tabular}
\quad
\begin{tabular}[b]{@{}p{0.425\textwidth}@{}}
\centering\includegraphics[width=1.2\linewidth]{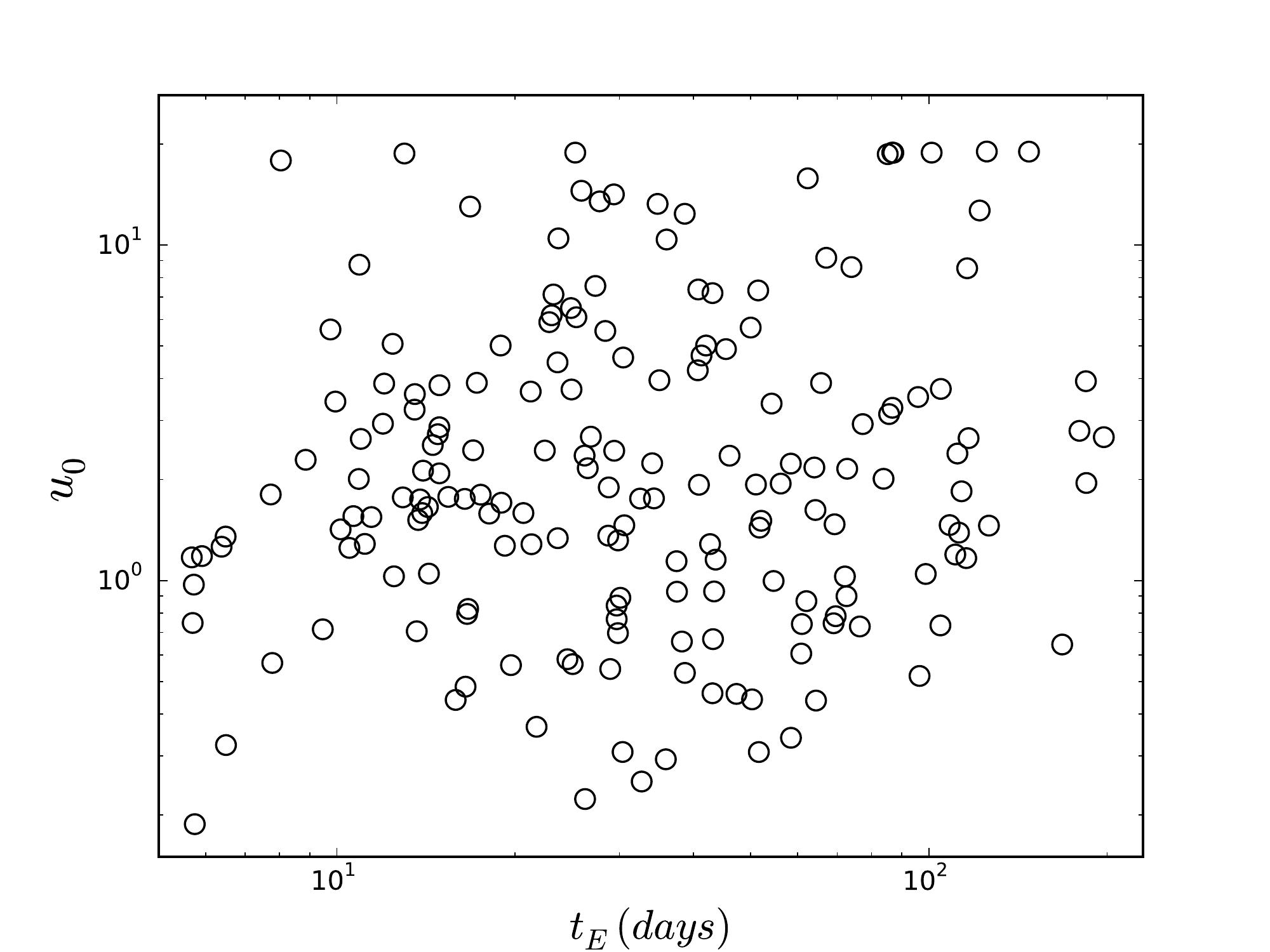} \\
\end{tabular} 
\caption{Left panel: timescale distribution of the complete sample  microlensing events (top histogram), compared with that of the RC subsample  (bottom histogram). The purple and cyan lines are the best Gaussian fits, with the mean  positions labelled. Right panel: distribution of the impact parameter $u_0$ and Einstein radius crossing time $t_E$ for the complete sample.}
\label{fig:timescale}
\end{figure}

The majority of the microlensing events in the sample region are expected to be bulge-bulge events and bulge-disk events (e.g. \citealt{Gould 95}), but at these latitudes there are also potentially disk-bulge events with the source in the far disk. Indeed, the foreground contamination by disk-disk events appears to be small, as we observe only half a dozen sources with blue-enough colors ($J-K_s \leq 2.0$) consistent with a foreground main-sequence disk population (\autoref{fig:cmd}).

With the information provided by the fitting procedure and the color-magnitude diagram, it is impossible to obtain all of the parameters needed to constrain the individual lens masses, except for the cases in which the parallax effects are evident.  As mentioned earlier, special events like parallax events will be analyzed in the future. However, for a large enough sample like ours, the distribution of timescales gives a global idea of mass distributions and tentative mass ranges that were detected (\autoref{fig:timescale}). \\

The shape of the timescale distribution is similar for the total sample and the RC sample. The peak of the timescale distribution, i.e., the most common value for the Einstein radius crossing time of the complete sample is $30.91\pm 1.08$ days, and for the RC sources is $29.93 \pm 1.06$ days. The RC sample mean is slightly shorter, but consistent within the errors. These mean values correspond to intermediate mass lenses (typical disk/bulge main-sequence stars) under reasonable model assumptions like those of the recent predictions of \cite{Wegg17}. The shape of the timescale distribution is also consistent with some previous studies in the bulge region \citep{lukaz15}. Both distributions follow a symmetric curve in $log(t_E)$, which is different, for example, from the distribution obtained by \cite{barry11}. This is probably due to the lack of short timescale VVV events. 

Both distributions are similar (\autoref{fig:timescale}), ranging from small values suggesting stellar mass objects to long duration events, which are generally associated with massive objects. Short timescale events with $t_E \leq 10$ days are lacking, and we argue that this is merely an effect of our low sampling efficiency for the short events in comparison with other surveys  like OGLE and MOA that have more frequent sampling and much longer timescale coverage. For example, the frequent sampling of the observations by \cite{Shvartzvald17} yield shorter timescale events in the mean (ranging from $t_E=7$ days to $30$ days). Their mean timescale, $t_E=17.2$ days, is significantly different than ours, and may suggest the presence of more massive lenses closer to the Galactic center or disk-disk events because of the low latitude of the studied area, but extreme caution is warranted with this comparison because of the different sample sizes and observing strategies.

On the other extreme of the timescale distribution, we observe a non-negligible number of long timescale events ($t_E \geq 100$ days) that are consistent with the presence of massive objects (in the black hole realm) or disk-disk events. However, as the value of the timescale is degenerate, it is necessary to do a more detailed study of these events, e.g. to include parallax in the fitting procedure and to model the inner Galaxy using different initial mass functions. These analyses are proposed for the future and are beyond the scope of this letter. 

Finally, the observed timescale and magnitude distribution of the detected events can be helpful to optimize the observational microlensing campaign of the WFIRST \citep{Spergel15}, and also to predict event rates and completeness. The observed magnitude ranges for the $J$ and $K_s$-bands 
($12 < J < 21.5$, and $11 < K_s < 17.5$, respectively),
and the color-magnitude diagrams show that the searches are more efficient at longer wavelengths. In fact, most of the photometric incompleteness in our sample is given by the lack of deeper J-band observations. \\


\section{Conclusions}
\label{sec:sec5}
For the first time, we have detected a large number of microlensing events around the Galactic center using the VVV near-IR photometry. 
We present the color-magnitude diagrams of the microlensing sources for the VVV tiles b332, b333, and b334, which show good qualitative agreement amongst themselves. There is an apparent excess of microlensing sources in the central tile b333 
in comparison with the average of the other two tiles, even though tile b333 is the most reddened and crowded of all.

We also presented the timescale distribution of the observed events that ranges from $5$ to $200$ days. We do not find significant numbers of events with $t_E<10$ days, due to our low-detection efficiency for short timescale events. There is, however, a non-negligible number of long timescale events ($t_E \geq 100$ days), which would be consistent with a population of massive black holes or disk-disk events.

This work demonstrates the usefulness of the VVV Survey to detect microlensing events in highly reddened and crowded areas like the Galactic center region. The present microlensing search covers the three most central VVV tiles, and can, in principle, be extended to adjacent areas that have not yet been studied due to heavy extinction. Such extended search would produce a complete timescale distribution map of the inner Milky Way bulge and show the dependencies with Galactic latitude and longitude, to complement previous bulge microlensing studies (e.g. \citealt{Popowski05}, \citealt{Sumi13}, \citealt{lukaz15}).

Our work also indicates that the microlensing optical depth keeps rising all the way to the Galactic center, but further observations are necessary to confirm this, and that a microlensing search in this region with the  WFIRST would be very profitable \citep{Spergel15}; and our results may be relevant to optimize the observational campaigns for that and other future surveys.  

\acknowledgments
We gratefully acknowledge data from the ESO Public Survey program ID 179.B-2002 taken with the VISTA telescope, and products from the Cambridge Astronomical Survey Unit (CASU). Support is provided by the BASAL Center for Astrophysics and Associated Technologies (CATA) through grant PFB-06, and the Ministry for the Economy, Development and Tourism, Programa Iniciativa Cient\'ifica Milenio grant IC120009, awarded to the Millennium Institute of Astrophysics (MAS). D.M. acknowledges support from FONDECYT regular grant No. 1170121. \\

\mbox{}

\clearpage
\begin{deluxetable}{cccccccccccc}
\tablecaption{VVV Survey first quality microlensing events data with their respective positions in equatorial coordinates, baseline $K_s$ magnitude, Color and the parameters obtained using the standard microlensing model including the blending ($f_{bl}$). The label RC correspond to the events located in the Red Clump and the O refers to the events that overlap. \label{tab:table1}}
\tablehead{
\colhead{Tile} & \colhead{ID} & \colhead{RA} & \colhead{DEC} & \colhead{$K_s$} & \colhead{$J-K$} & \colhead{Amp} & \colhead{$u_{0}$} & \colhead{$t_{0}$} & \colhead{$t_E$} & \colhead{$f_{bl}$} & Comment \\
 & & & & {\tiny (mag)} & {\tiny (mag)}  & &  & {\tiny (MJD)} & {\tiny (days)}  &
}
\startdata
b332 & 14-26290 & 265.09996 & -31.37107 & 15.56 & 3.28 & 1.47 & 0.45 & 56437.69 & 112.43 & 1.00 & RC \\
b332 & 14-55860 & 265.16328 & -31.41736 & 15.72 & 3.81 & 6.36 & 0.11 & 57243.97 & 40.78 & 1.00 & RC \\
b332 & 16-32398 & 264.84333 & -30.69719 & 15.44 & 4.31 & 3.60 & 0.22 & 56181.72 & 21.25 & 1.00 & RC \\
b332 & 18-36548 & 265.43515 & -30.96762 & 13.20 & 5.61 & 1.48 & 0.41 & 55792.95 & 13.93 & 1.00 &  \\
b332 & 18-41105 & 265.38645 & -31.05165 & 15.34 & 3.50 & 1.81 & 0.14 & 56478.57 & 72.58 & 0.15 & RC \\
b332 & 68-5694 & 265.48753 & -30.75808 & 15.30 & 3.70 & 0.68 & 0.48 & 55783.51 & 10.50 & 1.00 & RC \\
b332 & 68-14868 & 265.49243 & -30.78767 & 16.02 & 3.84 & 2.21 & 0.32 & 56487.96 & 34.08 & 1.00 & RC \\
b332 & 68-39711 & 265.50969 & -30.87205 & 12.51 & 6.72 & 1.72 & 0.38 & 56484.66 & 34.32 & 0.98 &  \\
b332 & 68-43156 & 265.48090 & -30.92659 & 16.64 & 2.76 & 8.24 & 0.05 & 56488.58 & 43.10 & 0.54 &  \\
b332 & 110-61443 & 265.06127 & -30.50046 & 15.13 & 3.69 & 1.07 & 0.05 & 56046.39 & 62.43 & 1.00 & RC \\
b332 & 110-74031 & 265.16668 & -30.40049 & 13.26 & 3.75 & 0.46 & 0.46 & 56552.93 & 64.47 & 0.33 & RC \\
b332 & 114-43783 & 265.32371 & -30.07666 & 14.59 & 3.65 & 0.79 & 0.65 & 56090.08 & 29.69 & 1.00 & RC \\
b332 & 21-40061 & 264.31254 & -30.77783 & 16.64 & 2.98 & 2.39 & 0.30 & 56558.71 & 16.99 & 1.00 &  \\
b332 & 23-82719 & 264.88706 & -31.20422 & 12.96 & 4.10 & 1.23 & 0.48 & 56895.29 & 19.22 & 1.00 &  \\
b332 & 27-8503 & 264.90537 & -30.87510 & 13.90 & 4.75 & 0.60 & 0.67 & 56202.48 & 12.00 & 0.55 & RC,O \\
b332 & 27-15159 & 264.99483 & -30.77833 & 13.94 & 4.59 & 2.02 & 0.25 & 56118.66 & 14.81 & 1.00 & RC \\
b332 & 27-31227 & 264.99579 & -30.84077 & 13.13 & 86.88 & 0.46 & 0.84 & 55799.69 & 16.49 & 1.00 &  \\
b332 & 27-38227 & 265.03919 & -30.80838 & 13.77 & 5.26 & 4.24 & 0.19 & 56874.75 & 40.70 & 0.96 &  \\
b332 & 34-1193 & 264.93179 & -31.15614 & 16.53 & 2.25 & 8.34 & 0.05 & 55792.34 & 116.06 & 0.45 &  \\
b332 & 34-4022 & 264.95451 & -31.13701 & 17.03 & 2.00 & 0.85 & 0.19 & 55783.65 & 16.44 & 0.48 &  \\
b332 & 34-57640 & 265.02460 & -31.25104 & 16.08 & 83.92 & 1.59 & 0.13 & 56163.10 & 56.19 & 0.28 &  \\
b332 & 38-16855 & 265.12081 & -30.96135 & 15.72 & 84.26 & 1.07 & 0.37 & 56491.21 & 21.31 & 0.71 &  \\
b332 & 44-42130 & 265.07155 & -31.13005 & 17.27 & 82.68 & 7.98 & 0.09 & 55799.68 & 36.05 & 1.00 &  \\
b332 & 48-5228 & 265.23875 & -30.75368 & 16.05 & 83.94 & 0.75 & 0.62 & 56107.66 & 29.69 & 1.00 &  \\
b332 & 48-81289 & 265.46393 & -30.74961 & 15.83 & 3.99 & 3.87 & 0.21 & 57249.02 & 65.75 & 1.00 & RC \\
b332 & 51-54921 & 264.45830 & -30.63581 & 16.77 & 2.17 & 1.41 & 0.05 & 56135.92 & 83.84 & 0.11 &  \\
b332 & 59-16446 & 264.73652 & -30.09485 & 15.72 & 1.84 & 1.47 & 0.41 & 56090.41 & 10.66 & 1.00 &  \\
b332 & 59-26357 & 264.79947 & -30.04314 & 12.87 & 2.15 & 9.87 & 0.05 & 56196.62 & 23.67 & 0.55 & RC \\
b332 & 66-77202 & 265.00700 & -30.63918 & 14.98 & 3.89 & 0.53 & 0.81 & 55812.01 & 96.44 & 1.00 & RC \\
b332 & 211-37537 & 265.29213 & -30.44556 & 11.48 & 4.16 & 0.46 & 0.36 & 55833.97 & 12.92 & 1.00 &  \\
b332 & 213-59426 & 264.96048 & -29.96142 & 13.71 & 2.15 & 0.33 & 1.00 & 56484.12 & 51.61 & 0.90 & RC \\
b332 & 213-72294 & 264.97054 & -29.99633 & 13.12 & 2.71 & 0.66 & 0.71 & 56149.62 & 43.19 & 1.00 & RC \\
b332 & 310-80902 & 264.98350 & -30.33055 & 16.95 & 2.33 & 1.35 & 0.05 & 55813.77 & 113.55 & 0.10 &  \\
b332 & 310-100917 & 265.05852 & -30.29863 & 15.82 & 84.23 & 1.78 & 0.05 & 56517.63 & 104.76 & 0.20 &  \\
b332 & 312-5320 & 265.30551 & -30.66024 & 16.36 & 83.29 & 6.39 & 0.05 & 55787.84 & 25.27 & 1.00 &  \\
b332 & 410-53003 & 265.00924 & -30.19381 & 14.43 & 2.94 & 0.86 & 0.55 & 57202.94 & 72.14 & 1.00 & RC \\
b332 & 414-8102 & 265.14228 & -29.85276 & 13.76 & 2.45 & 0.39 & 0.49 & 56200.05 & 15.87 & 0.36 & RC \\
b332 & 414-18576 & 265.15753 & -29.87074 & 13.66 & 86.34 & 1.57 & 0.09 & 55812.30 & 22.84 & 0.61 &  \\
b332 & 414-69038 & 265.26596 & -29.91121 & 15.92 & 3.36 & 2.42 & 0.28 & 56495.48 & 26.85 & 1.00 &  \\
b332 & 511-90196 & 265.46836 & -30.39392 & 15.24 & 4.58 & 1.55 & 0.33 & 56062.48 & 72.79 & 1.00 & RC \\
b332 & 513-61175 & 265.06202 & -29.82790 & 16.24 & 2.33 & 3.23 & 0.28 & 56120.77 & 116.70 & 1.00 &  \\
b332 & 515-19885 & 265.52832 & -30.05048 & 13.80 & 2.41 & 0.70 & 0.20 & 55790.66 & 29.83 & 0.17 & RC \\
b332 & 610-23832 & 265.17869 & -30.19984 & 13.64 & 3.28 & 0.24 & 0.31 & 57271.00 & 25.02 & 0.25 & RC \\
b332 & 610-47014 & 265.21751 & -30.22933 & 13.48 & 2.85 & 0.32 & 1.00 & 56085.32 & 6.49 & 0.95 & RC \\
b332 & 610-78499 & 265.26797 & -30.27310 & 12.92 & 3.13 & 0.65 & 0.66 & 56451.52 & 69.02 & 1.00 & RC \\
b332 & 610-97533 & 265.29367 & -30.30727 & 13.22 & 3.77 & 0.29 & 1.00 & 56059.03 & 58.48 & 0.99 & RC \\
b332 & 614-45372 & 265.35913 & -30.03531 & 12.38 & 4.11 & 3.89 & 0.21 & 56732.81 & 91.71 & 1.00 & O \\
b332 & 616-24844 & 265.94883 & -30.18384 & 15.19 & 2.72 & 1.35 & 0.05 & 57239.33 & 77.31 & 0.17 & RC \\
b333 & 12-42539 & 265.50396 & -29.81472 & 16.76 & 83.01 & 1.75 & 0.40 & 56124.13 & 14.24 & 1.00 &  \\
b333 & 12-69304 & 265.52124 & -29.88285 & 14.33 & 3.72 & 0.81 & 0.18 & 56825.95 & 76.48 & 0.16 & RC \\
b333 & 14-55156 & 266.02807 & -30.18754 & 16.74 & 83.32 & 4.38 & 0.05 & 56015.03 & 85.21 & 1.00 &  \\
b333 & 23-1263 & 265.57612 & -29.91941 & 14.86 & 4.49 & 7.37 & 0.05 & 56036.53 & 23.22 & 0.39 & RC \\
b333 & 23-4699 & 265.66126 & -29.81254 & 13.17 & 2.71 & 0.41 & 0.65 & 56123.65 & 21.74 & 0.48 & RC \\
b333 & 27-90082 & 265.96369 & -29.68600 & 14.35 & 5.36 & 0.80 & 0.30 & 57248.76 & 30.11 & 0.36 & RC \\
b333 & 211-1492 & 266.00663 & -29.31280 & 14.73 & 4.72 & 4.63 & 0.05 & 56038.23 & 23.05 & 0.33 & RC \\
b333 & 211-15629 & 266.08567 & -29.24704 & 12.31 & 4.71 & 1.40 & 0.42 & 56123.94 & 28.73 & 0.90 &  \\
b333 & 211-17696 & 266.05926 & -29.29131 & 17.01 & 4.35 & 4.11 & 0.05 & 56101.05 & 25.88 & 1.00 &  \\
b333 & 213-8425 & 265.80896 & -28.57501 & 14.13 & 3.28 & 3.07 & 0.11 & 57251.15 & 42.03 & 0.67 & RC \\
b333 & 215-18630 & 266.33176 & -28.90853 & 14.83 & 85.15 & 4.41 & 0.18 & 55998.65 & 23.59 & 1.00 &  \\
b333 & 16-86247 & 265.76515 & -29.60036 & 17.13 & 82.82 & 1.64 & 0.28 & 56838.36 & 10.89 & 1.00 &  \\
b333 & 21-1306 & 265.02993 & -29.64241 & 12.73 & 2.05 & 1.66 & 0.37 & 56037.80 & 7.73 & 1.00 &  \\
b333 & 21-106713 & 265.21343 & -29.75273 & 15.37 & 1.50 & 7.18 & 0.05 & 56019.43 & 125.27 & 1.00 &  \\
b333 & 25-1139 & 265.33053 & -29.21758 & 15.03 & 1.95 & 5.17 & 0.05 & 56036.98 & 10.91 & 1.00 &  \\
b333 & 25-13345 & 265.28460 & -29.32971 & 14.17 & 2.12 & 6.51 & 0.05 & 57279.95 & 34.81 & 0.95 & RC \\
b333 & 25-69718 & 265.43930 & -29.32309 & 15.60 & 4.25 & 1.60 & 0.08 & 56864.43 & 111.72 & 0.22 & RC \\
b333 & 29-89508 & 265.63252 & -29.09897 & 15.33 & 4.41 & 1.96 & 0.20 & 56373.39 & 8.85 & 0.79 & RC \\
b333 & 110-14235 & 265.92726 & -29.11678 & 14.80 & 4.63 & 1.70 & 0.39 & 56024.59 & 18.96 & 1.00 & RC \\
b333 & 110-84253 & 266.00399 & -29.24772 & 12.38 & 4.87 & 0.66 & 0.56 & 56823.78 & 5.71 & 1.00 &  \\
b333 & 110-103868 & 266.02570 & -29.28402 & 13.30 & 5.48 & 1.16 & 0.31 & 55812.12 & 14.90 & 1.00 &  \\
b333 & 513-21168 & 265.92994 & -28.44640 & 14.04 & 2.44 & 3.43 & 0.14 & 56833.34 & 9.94 & 0.53 & RC \\
b333 & 18-90199 & 266.33387 & -29.88208 & 14.42 & 85.57 & 1.00 & 0.39 & 56175.52 & 13.72 & 1.00 &  \\
b333 & 32-17832 & 265.34470 & -29.60216 & 14.52 & 85.46 & 6.09 & 0.13 & 56037.96 & 24.85 & 1.00 &  \\
b333 & 32-64434 & 265.36278 & -29.73430 & 15.41 & 3.36 & 1.92 & 0.28 & 57170.22 & 13.98 & 0.81 & RC \\
b333 & 32-65431 & 265.42744 & -29.64701 & 13.14 & 3.39 & 1.97 & 0.09 & 56029.38 & 18.91 & 0.72 & RC \\
b333 & 32-75883 & 265.42179 & -29.69092 & 14.30 & 3.84 & 2.41 & 0.29 & 56136.14 & 22.44 & 0.96 & RC \\
b333 & 34-42751 & 265.85305 & -30.03220 & 14.46 & 85.56 & 1.19 & 0.05 & 56474.62 & 115.65 & 0.06 &  \\
b333 & 34-49917 & 265.94819 & -29.92529 & 16.42 & 83.49 & 7.44 & 0.05 & 56017.23 & 147.62 & 1.00 &  \\
b333 & 34-81174 & 265.94033 & -30.05182 & 12.97 & 4.07 & 1.01 & 0.56 & 56032.42 & 54.66 & 1.00 &  \\
b333 & 36-26139 & 265.52159 & -29.38604 & 17.23 & 2.75 & 2.85 & 0.17 & 56038.73 & 11.96 & 0.60 &  \\
b333 & 36-36965 & 265.58457 & -29.33481 & 14.98 & 3.71 & 6.46 & 0.05 & 56509.76 & 27.33 & 0.71 & RC \\
b333 & 42-1507 & 265.35617 & -29.53228 & 11.47 & 3.51 & 1.31 & 0.17 & 56841.00 & 22.34 & 1.00 & O \\
b333 & 42-4246 & 265.37832 & -29.51096 & 11.57 & 3.48 & 0.60 & 0.16 & 56085.67 & 60.87 & 0.12 &  \\
b333 & 44-97159 & 266.06948 & -29.93123 & 14.02 & 4.98 & 0.66 & 0.05 & 56893.72 & 12.66 & 1.00 & RC, O \\
b333 & 46-40910 & 265.71181 & -29.17016 & 14.83 & 4.46 & 1.33 & 0.36 & 56060.26 & 51.03 & 1.00 & RC \\
b333 & 51-33207 & 265.26581 & -29.42843 & 13.42 & 3.03 & 2.60 & 0.20 & 56144.52 & 10.97 & 0.65 & RC \\
b333 & 53-62793 & 265.84453 & -29.75693 & 14.23 & 1.97 & 0.49 & 0.61 & 56417.41 & 62.10 & 1.00 &  \\
b333 & 57-41448 & 266.02671 & -29.42235 & 13.67 & 5.70 & 0.78 & 0.54 & 56821.26 & 14.30 & 1.00 &  \\
b333 & 59-11271 & 265.64080 & -28.81314 & 15.95 & 3.03 & 3.10 & 0.23 & 57249.42 & 54.24 & 1.00 &  \\
b333 & 62-70519 & 265.72288 & -29.59623 & 16.97 & 82.98 & 5.58 & 0.12 & 56019.40 & 51.48 & 1.00 &  \\
b333 & 64-31424 & 266.18966 & -29.86641 & 14.64 & 3.59 & 0.72 & 0.68 & 56094.53 & 13.64 & 1.00 & RC \\
b333 & 64-49508 & 266.15813 & -29.97755 & 15.35 & 84.70 & 4.62 & 0.05 & 56071.79 & 121.84 & 0.67 &  \\
b333 & 64-96023 & 266.29007 & -29.97366 & 14.67 & 4.19 & 5.40 & 0.14 & 56184.72 & 25.38 & 1.00 & RC \\
b333 & 66-21828 & 265.78826 & -29.34844 & 16.30 & 83.68 & 3.29 & 0.24 & 55780.57 & 86.78 & 1.00 &  \\
b333 & 66-52469 & 265.85955 & -29.35649 & 14.71 & 4.41 & 1.69 & 0.36 & 56857.92 & 40.88 & 1.00 & RC \\
b333 & 68-104841 & 266.51787 & -29.66045 & 14.31 & 4.72 & 4.07 & 0.05 & 56506.52 & 27.79 & 1.00 & RC \\
b333 & 112-106497 & 266.65154 & -29.47018 & 14.78 & 4.29 & 3.24 & 0.05 & 56566.06 & 29.37 & 1.00 & RC \\
b333 & 114-13230 & 266.08843 & -28.88644 & 15.08 & 4.36 & 4.14 & 0.09 & 56490.79 & 50.01 & 0.54 & RC \\
b333 & 114-46792 & 266.22084 & -28.81652 & 10.72 & 4.59 & 1.24 & 0.46 & 56811.23 & 42.69 & 0.97 &  \\
b333 & 116-37815 & 266.66043 & -29.21073 & 13.77 & 4.58 & 1.07 & 0.44 & 56099.16 & 10.14 & 1.00 & RC \\
b333 & 116-79669 & 266.81518 & -29.13610 & 12.81 & 4.72 & 0.22 & 1.00 & 56061.13 & 32.72 & 0.74 &  \\
b333 & 310-29281 & 265.78988 & -29.01572 & 13.91 & 3.97 & 0.67 & 0.16 & 56151.46 & 11.15 & 0.25 & RC \\
b333 & 310-79969 & 265.80993 & -29.16381 & 13.83 & 3.82 & 1.32 & 0.05 & 56042.08 & 12.42 & 0.39 & RC \\
b333 & 310-96589 & 265.82793 & -29.19609 & 13.44 & 4.32 & 1.55 & 0.41 & 56044.91 & 18.08 & 1.00 &  \\
b333 & 312-2696 & 266.19753 & -29.39797 & 16.01 & 83.86 & 3.76 & 0.05 & 56010.00 & 86.72 & 1.00 &  \\
b333 & 312-44070 & 266.28416 & -29.41028 & 12.69 & 87.31 & 1.06 & 0.55 & 56375.82 & 12.48 & 1.00 &  \\
b333 & 312-62428 & 266.36750 & -29.35349 & 12.52 & 6.83 & 0.95 & 0.10 & 56925.41 & 108.42 & 0.16 &  \\
b333 & 314-29796 & 265.94244 & -28.80337 & 14.66 & 3.66 & 0.78 & 0.64 & 56384.91 & 69.60 & 1.00 & RC \\
b333 & 314-55699 & 266.06406 & -28.72311 & 15.59 & 3.89 & 2.00 & 0.05 & 56493.50 & 13.00 & 1.00 & RC \\
b333 & 314-77236 & 266.03154 & -28.84794 & 13.92 & 3.94 & 0.21 & 0.67 & 56160.58 & 26.26 & 0.30 & RC \\
b333 & 316-15279 & 266.48065 & -29.03126 & 14.83 & 4.67 & 0.82 & 0.05 & 56189.99 & 8.04 & 1.00 & RC \\
b333 & 316-40378 & 266.53643 & -29.03117 & 13.76 & 5.29 & 0.73 & 0.67 & 56811.65 & 61.01 & 1.00 &  \\
b333 & 316-90842 & 266.62655 & -29.07436 & 12.75 & 5.38 & 2.38 & 0.19 & 56813.44 & 14.90 & 0.83 &  \\
b333 & 316-100780 & 266.62076 & -29.11919 & 14.58 & 2.10 & 1.71 & 0.41 & 56125.07 & 20.65 & 1.00 & RC \\
b333 & 410-7976 & 265.84798 & -28.85905 & 11.51 & 5.12 & 1.22 & 0.47 & 55801.43 & 29.85 & 1.00 &  \\
b333 & 412-26866 & 266.43900 & -29.12995 & 12.49 & 5.17 & 1.49 & 0.44 & 57225.16 & 126.32 & 1.00 &  \\
b333 & 414-25228 & 266.02719 & -28.66408 & 15.22 & 3.78 & 3.74 & 0.05 & 56525.74 & 67.09 & 0.51 & RC \\
b333 & 414-49757 & 266.09186 & -28.65845 & 13.23 & 86.78 & 1.74 & 0.29 & 56087.28 & 32.52 & 0.70 &  \\
b333 & 416-65395 & 266.66200 & -28.93446 & 13.07 & 86.94 & 0.57 & 0.76 & 56159.63 & 24.51 & 1.00 &  \\
b333 & 515-29074 & 266.46354 & -28.75120 & 14.39 & 85.59 & 1.40 & 0.44 & 57276.83 & 51.74 & 1.00 &  \\
b333 & 515-45486 & 266.51525 & -28.73049 & 13.48 & 3.26 & 0.81 & 0.58 & 57187.48 & 43.37 & 1.00 & RC \\
b333 & 515-49289 & 266.46807 & -28.81187 & 12.65 & 4.48 & 1.33 & 0.35 & 57114.57 & 184.52 & 1.00 &  \\
b333 & 610-11467 & 265.98079 & -29.03053 & 12.35 & 4.53 & 0.32 & 0.89 & 56514.01 & 30.38 & 0.71 &  \\
b333 & 610-36012 & 266.06155 & -28.99876 & 12.33 & 4.41 & 0.22 & 0.80 & 56417.64 & 38.71 & 1.00 &  \\
b333 & 610-40425 & 266.08328 & -28.98288 & 13.74 & 4.28 & 3.07 & 0.21 & 56138.31 & 13.54 & 0.96 & RC \\
b333 & 610-50685 & 266.12013 & -28.96453 & 13.32 & 4.80 & 1.84 & 0.36 & 56114.15 & 28.78 & 1.00 &  \\
b333 & 612-80860 & 266.71318 & -29.28126 & 14.95 & 4.97 & 1.18 & 0.50 & 56035.82 & 5.69 & 1.00 & RC \\
b333 & 612-98993 & 266.67770 & -29.40080 & 12.69 & 4.45 & 0.88 & 0.33 & 57261.87 & 37.54 & 0.42 &  \\
b333 & 614-70531 & 266.34156 & -28.72281 & 14.24 & 4.94 & 0.47 & 0.65 & 56095.26 & 50.27 & 0.57 & RC \\
b333 & 614-84336 & 266.35631 & -28.74884 & 14.35 & 4.61 & 6.74 & 0.05 & 57260.35 & 38.69 & 1.00 & RC \\
b333 & 614-94658 & 266.41497 & -28.69935 & 12.85 & 4.16 & 1.49 & 0.26 & 56121.04 & 6.49 & 0.97 &  \\
b334 & 16-74959 & 266.61660 & -28.34858 & 14.75 & 5.69 & 2.39 & 0.24 & 56128.88 & 26.20 & 0.72 & RC \\
b334 & 21-23644 & 265.97740 & -28.38942 & 14.50 & 3.24 & 1.54 & 0.42 & 55820.68 & 11.43 & 1.00 & RC \\
b334 & 34-25019 & 266.68395 & -28.78282 & 14.66 & 5.83 & 1.66 & 0.40 & 56173.57 & 64.33 & 1.00 & RC \\
b334 & 34-82386 & 266.88664 & -28.72094 & 12.37 & 6.28 & 1.67 & 0.00 & 55992.50 & 47.02 & 0.20 &  \\
b334 & 36-13718 & 266.32706 & -28.19646 & 14.46 & 2.39 & 3.59 & 0.17 & 57251.46 & 45.42 & 1.00 & RC \\
b334 & 36-20307 & 266.40409 & -28.10910 & 13.34 & 3.78 & 1.05 & 0.38 & 56856.98 & 15.42 & 1.00 & RC \\
b334 & 36-83203 & 266.50899 & -28.18965 & 17.28 & 2.53 & 13.40 & 0.02 & 56126.77 & 31.55 & 0.35 &  \\
b334 & 38-32772 & 266.99787 & -28.35616 & 14.26 & 3.89 & 0.95 & 0.00 & 56009.09 & 77.89 & 0.10 & RC \\
b334 & 42-83526 & 266.34657 & -28.42009 & 14.53 & 4.08 & 2.23 & 0.32 & 57245.15 & 58.45 & 1.00 & RC \\
b334 & 44-14992 & 266.84899 & -28.50806 & 15.40 & 84.55 & 2.56 & 0.29 & 56182.30 & 14.52 & 1.00 &  \\
b334 & 44-77077 & 266.92793 & -28.63633 & 14.24 & 4.79 & 0.79 & 0.26 & 56837.94 & 104.58 & 0.25 & RC \\
b334 & 110-47231 & 266.84039 & -27.93049 & 15.85 & 4.14 & 4.86 & 0.15 & 56842.88 & 9.75 & 1.00 & RC \\
b334 & 110-81842 & 266.91081 & -27.95946 & 14.40 & 4.42 & 1.12 & 0.13 & 56004.80 & 110.82 & 0.18 & RC \\
b334 & 112-80046 & 267.40998 & -28.27695 & 14.39 & 4.65 & 2.84 & 0.27 & 55807.07 & 179.63 & 1.00 & RC \\
b334 & 112-91317 & 267.39564 & -28.33975 & 11.11 & 3.79 & 1.47 & 0.43 & 56583.86 & 33.65 & 1.00 & O \\
b334 & 616-64986 & 267.71918 & -27.77335 & 14.67 & 3.06 & 1.38 & 0.35 & 56516.10 & 13.82 & 1.00 & RC \\
b334 & 14-35957 & 266.91087 & -28.86083 & 14.89 & 4.25 & 1.05 & 0.00 & 56266.64 & 101.01 & 1.00 & RC \\
b334 & 114-2993 & 266.95813 & -27.59214 & 14.23 & 3.81 & 0.85 & 0.05 & 56897.19 & 24.42 & 1.00 & RC,O \\
b334 & 114-34851 & 266.99985 & -27.65723 & 12.90 & 5.39 & 3.97 & 0.18 & 56099.08 & 30.45 & 1.00 &  \\
b334 & 114-86000 & 267.15587 & -27.65442 & 11.40 & 5.52 & 0.28 & 1.00 & 56559.30 & 35.94 & 0.86 &  \\
b334 & 116-26287 & 267.55340 & -27.87324 & 13.52 & 3.85 & 0.96 & 0.45 & 56487.61 & 5.73 & 0.71 & RC \\
b334 & 116-34279 & 267.50475 & -27.97748 & 11.93 & 0.44 & 0.53 & 0.78 & 56003.36 & 19.68 & 1.00 &  \\
b334 & 116-80676 & 267.64681 & -27.95567 & 14.34 & 3.73 & 5.29 & 0.15 & 56158.40 & 28.41 & 1.00 & RC \\
b334 & 316-42169 & 267.36471 & -27.86155 & 16.13 & 3.56 & 2.90 & 0.19 & 57170.53 & 16.68 & 0.73 &  \\
b334 & 316-51711 & 267.40790 & -27.83484 & 14.14 & 4.98 & 0.70 & 0.68 & 56110.12 & 9.46 & 1.00 & RC \\
b334 & 316-90309 & 267.49660 & -27.85982 & 15.09 & 4.62 & 1.55 & 0.20 & 56190.74 & 12.02 & 1.00 & RC \\
b334 & 23-24739 & 266.51787 & -28.66503 & 13.42 & 5.56 & 0.67 & 0.21 & 56472.64 & 167.96 & 0.17 &  \\
b334 & 23-41942 & 266.51526 & -28.73046 & 13.48 & 3.25 & 0.75 & 0.16 & 57188.50 & 98.78 & 0.19 & RC \\
b334 & 25-71070 & 266.25801 & -28.15896 & 12.57 & 3.31 & 0.65 & 0.71 & 55806.34 & 38.26 & 1.00 &  \\
b334 & 51-10132 & 266.06642 & -28.21192 & 15.15 & 2.34 & 1.84 & 0.00 & 56499.51 & 17.47 & 0.15 & RC \\
b334 & 51-56759 & 266.18724 & -28.21185 & 15.69 & 2.98 & 6.96 & 0.06 & 56099.35 & 73.99 & 0.56 &  \\
b334 & 51-85618 & 266.18884 & -28.31620 & 14.30 & 2.69 & 2.43 & 0.29 & 56139.42 & 29.41 & 0.97 & RC \\
b334 & 55-55967 & 266.40621 & -27.87824 & 14.37 & 2.81 & 2.11 & 0.10 & 57233.42 & 35.06 & 0.49 & RC \\
b334 & 55-58674 & 266.37368 & -27.93529 & 14.77 & 2.36 & 6.44 & 0.04 & 56198.53 & 16.91 & 0.69 & RC \\
b334 & 64-18928 & 267.04305 & -28.59909 & 14.32 & 5.66 & 0.97 & 0.28 & 57071.44 & 197.48 & 1.00 & RC \\
b334 & 64-19410 & 267.01037 & -28.64823 & 16.30 & 2.34 & 3.26 & 0.25 & 56174.12 & 85.67 & 1.00 &  \\
b334 & 68-52576 & 267.21228 & -28.47343 & 16.89 & 2.48 & 10.44 & 0.00 & 57252.55 & 82.48 & 1.00 &  \\
b334 & 211-31387 & 266.93040 & -28.09790 & 13.51 & 7.44 & 0.81 & 0.64 & 56196.01 & 16.59 & 1.00 &  \\
b334 & 213-31629 & 266.64126 & -27.46364 & 14.53 & 1.51 & 0.48 & 0.55 & 56158.10 & 43.10 & 0.45 &  \\
b334 & 310-31814 & 266.59112 & -27.87853 & 15.75 & 4.28 & 1.08 & 0.51 & 56001.37 & 37.48 & 1.00 & RC \\
b334 & 412-43006 & 267.33083 & -27.91374 & 15.44 & 4.14 & 1.34 & 0.09 & 56203.39 & 46.06 & 0.26 & RC \\
b334 & 412-79563 & 267.35098 & -28.02180 & 14.56 & 85.43 & 0.57 & 0.49 & 56836.85 & 5.91 & 1.00 &  \\
b334 & 414-10765 & 266.95702 & -27.26925 & 13.66 & 3.39 & 0.43 & 0.62 & 56519.89 & 47.31 & 0.56 & RC \\
b334 & 416-77898 & 267.63206 & -27.62068 & 15.10 & 4.73 & 0.78 & 0.63 & 56120.46 & 16.66 & 1.00 & RC \\
b334 & 511-36360 & 267.11900 & -27.83605 & 11.39 & 5.13 & 0.18 & 0.02 & 56373.36 & 153.28 & 0.00 &  \\
b334 & 513-35349 & 266.79474 & -27.25379 & 13.99 & 2.55 & 0.42 & 0.77 & 56837.28 & 7.78 & 1.00 & RC \\
b334 & 515-58251 & 267.30222 & -27.64802 & 14.27 & 85.74 & 0.79 & 0.41 & 57203.23 & 52.10 & 0.94 &  \\
b334 & 612-90493 & 267.57944 & -28.07189 & 15.32 & 4.25 & 1.48 & 0.25 & 55818.58 & 69.29 & 0.47 & RC \\
b334 & 614-30427 & 267.15078 & -27.41083 & 15.54 & 5.20 & 3.48 & 0.17 & 57261.64 & 95.89 & 0.72 & RC \\
b334 & 614-30702 & 267.12751 & -27.44610 & 13.13 & 3.54 & 1.24 & 0.37 & 56832.55 & 6.38 & 0.68 & RC \\
b334 & 614-71749 & 267.20863 & -27.48013 & 14.66 & 3.33 & 1.14 & 0.51 & 57251.94 & 43.63 & 1.00 & RC \\
b334 & 12-66675 & 266.43853 & -28.57618 & 12.62 & 4.49 & 1.71 & 0.32 & 56121.19 & 17.49 & 0.80 &  \\
\enddata
\end{deluxetable}

\end{document}